\documentclass[%
 reprint,
 amsmath,amssymb,
 aps,
]{revtex4-2}

\usepackage{graphicx}
\usepackage{dcolumn}
\usepackage{bm}

\begin{document}

\preprint{APS/123-QED}

\title{Compact Q-Stars with q,p-Deformed Fermions }

\author{Emre Dil}
\email{emredil@beylent.edu.tr}
 \affiliation{Beykent University, Engineering Faculty,\\ Ayazaga, Sariyer, Istanbul-Turkey}
\author{Ahmet Mecit Öztaş}%
 \email{oztas@hacettepe.edu.tr}
\affiliation{Hacettepe University, Dept. of Physics Engineering,\\ Beytepe, Cankaya, Ankara-Turkey}%

\date{\today}

\begin{abstract}
The Q-Star structures, also called as gray holes with a radius of two times greater than the corresponding Schwarzschild radius of same mass and from the compact neutron star family with a very high density, have been obtained by using a generalization of $q$-deformed fermions with two parameters $q$ and $p$ as the constituents of the star. Because the interaction between particles are controlled by deformation parameters in $q,p$-fermions instead of a complicated interaction Lagrangian, we consider the deformation parameters giving the maximum Q-Star pressure between the interacting particles of stars. The cold and hot Q-Stars in temperatures of $T=0$, $T=40\ \mathrm{MeV}$ and $T=60\ \mathrm{MeV}$ have been investigated from the numerical solutions of TOV equations and it is obtained that the Q-Stars can reach up to $13$ solar mass and $75$ Km radius. Moreover, it is found that the total mass and radius of the star increase when the star gets cool down, and a much denser state is reached.

\end{abstract}

\keywords{Compact stars; Q-Stars; Neutron stars, Structure and stability; Deformed fermions; Deformed Fermi gas model}
\maketitle


\section{\label{sec1}Introduction}

Understanding the interaction nature of particles system and their internal structures is very crucial in many-body quantum studies. The Lie groups and their corresponding algebras as symmetry transformations are well known in physics which cannot be used in complicated problems, instead we require more generalized symmetry concepts. An effective method to describe these complex systems is to consider a generalized particle model which is described by recently discovered quantum group symmetries which is led by the theory of quantum integrable systems \cite{azmi2006}. Quantum groups and quantum algebras, such as $SU_q(n)$, are specific deformations of the underlying usual Lie groups and Lie algebras by some deformation parameter q which results as a $q$-deformed boson or $q$-deformed fermion oscillator algebras, and their generalized versions with two parameters \cite{algin2005,jimbo1986,yu1989,woro1987}. Although the particle algebras are deformed it is obtained that the deformed algebras have still quantum group symmetry. Therefore, the existing quantum groups leaving particle algebras invariant, whose application area is quantum field theory are crucial in investigation of complicated interactions of many-body quantum systems.

The use of deformed fermions and bosons has become widespread in non-linear phenomena and much complicated areas of theoretical and experimental physics \cite{6,7,8}, such as in $N=2$ supersymmetry (SUSY) algebras \cite{9,10}, solid state physics \cite{11}, black hole physics \cite{12,13,14}, and in statistical mechanics \cite{15,16}. The use of quantum group particles makes more complex interactions between particles simplified by controlling the deformation parameters to describe the interaction between particles, instead of determining complicated interaction Hamiltonians. For example, one-parameter ($q$) and two-parameter ($q,p$) deformed bosons are considered to investigate the effective interactions between the particles and their resulting composite structures \cite{17,18,19,20}. Similar to those applications of deformed bosons, one-parameter and two-paramater deformed fermion counterparts have also been introduced to describe the complicated interactions by deformation parameters  \cite{15,21,22,23,24,25,26,27,28}. The use of two deformation parameters $q,p$ is important due to the following reasons; they are the most general oscillators having $SU_{q/p}(n)$ quantum group invariance \cite{29,30} and it enables to describe the phase transitions between bosonic and fermionic phases by adjusting the deformation parameters \cite{31,32}, which cannot be described by both one parameter deformed $SU_q(2)$ fermion model \cite{33,34} and standard undeformed fermion model \cite{35}. As a third advantage of using two parameter deformation of fermions is to provide more flexibility to explain some non-linear phenomena especially in phenomenological studies. For instance, some critical values of the two deformation parameters have been observed to fit the real laser properties \cite{36}.

In addition to the above applications of two-parameter fermions, there are some other applications related to our study on compact Q-Star. The investigation of some hadronic properties as the dynamical mass generation for quarks and the nuclear pairing force of the Bardeen–Cooper–Schrieffer (BCS) many-body formalism \cite{37,38} are fundamental interactions in the core of neutron stars. Another direct application of deformed particle algebras which is related to one of the special type of compact stars, is on preons as the constituents of compact stars. The preons could also be described by deformed particle algebras having an $SL_q(2)$ quantum group invariance \cite{39,40}. These studies lead us to ask whether the deformed fermions can be the constituents of a family of compact stars.

In astrophysics studies, Q-stars are described as the stars having a very high density and mass, and a radius about 2 times greater than the Schwarzschild radius of the corresponding mass. They are categorized in the most dense spectrum of the neutron stars with exotic constituents such as SUSY Q-balls \cite{41,42,42a}. Because of their very high density they are also called as the gray holes, meaning very close to form a black hole. The studies for describing the internal structures of Q-Stars have been attempted with fermions \cite{43}, and obtained very complicated interaction Lagrangians. This motivated us to investigate whether the internal structure of these Q-Stars can be described by $q,p$-deformed fermions which can control these complicated interactions by the deformation parameters.

Therefore, we first use the thermodynamical properties of $q,p$-deformed fermions called as Fermionic Fibonacci oscillators. In order to solve Tolman–-Oppenheimer-–Volkoff (TOV) equations for our star formed by $q,p$-fermions, we obtain the pressure, total number density and energy density functions for high densities, as neutron stars, much larger than the nuclear density. Then, we solve the TOV equations numerically to obtain the mass-radius relations of our $q,p$ star model. The results are consistent with Q-Star configurations. Also, we reach the consistent and similar results with the fermion Q-Stars results. We choose the parameters giving maximum pressure for the related number density. The resulting mass-radius relations are much beyond the neutron star structure and are consistent with Q-Stars because the $q,p$-fermions, as stated in above studies allows a transition from fermionic to bosonic phase and construction of BCS superconducting phases as in quark stars by controlling the deformation parameters.

\section{\label{sec2}Thermodynamics of Q-Stars with  $q,p$-Deformed Fermions}

Two-parameter $q,p$-deformed fermions are described by the following algebra between the creation and annihilation operators $\psi^{\dagger}$ and $\psi$, respectively \cite{31}
\begin{eqnarray}
    \label{1a}
    \psi_i\psi_j^{\dagger}+qp\psi_j^{\dagger}\psi_i&=&0, \qquad i\neq j \nonumber \\
    \psi_i\psi_j+\frac{q}{p}\psi_j^{\dagger}\psi_i&=&0, \qquad i<j \nonumber \\
    \psi_i^2&=&0, \nonumber \\
    \psi_1\psi_1^{\dagger}+p^2\psi_1^{\dagger}\psi_1&=&p^{2\hat{N}}, \nonumber \\
    \psi_i\psi_i^{\dagger}+q^2\psi_i^{\dagger}\psi_i&=&\psi_{i+1}\psi_{i+1}^{\dagger}+p^2\psi_j^{\dagger}\psi_i, \ i=1,2,...,d-1, \nonumber \\
    \psi_d\psi_d^{\dagger}+q^2\psi_d^{\dagger}\psi_d&=&q^{2\hat{N}},
\end{eqnarray}
where $\hat{N}$ is the number operator. The deformed Fermi-Dirac distribution function $\eta(\epsilon,q,p)$ is given as
\begin{equation}
    \label{4}
    \eta(\epsilon,q,p)=\frac{1}{\left|\ln(q^2/p^2)\right|}\left| \ln{\frac{e^{(\epsilon-\mu)/T}+q^2}{e^{(\epsilon-\mu)/T}+q^2}}\right|, \\
\end{equation}
where $T$ is temperature, $\epsilon$ is the single state energy and $\mu$ is chemical potential of $q,p$-fermions. Then, the total number of particles and energy for $q,p$-fermion system are obtained from \cite{16}
\begin{equation}
    \label{1b}
    N=\int^{\infty}_{0}{\eta(\epsilon,q,p)C\sqrt{\epsilon}d\epsilon},
\end{equation}
and
\begin{equation}
    \label{1c}
    U=\int^{\infty}_{0}{\epsilon\eta(\epsilon,q,p)C\sqrt{\epsilon}d\epsilon},
\end{equation}
where $C=(V/2\pi^2)(8\pi^2m)^{3/2}$. After solving these equation according to the $z=e^{\mu/T}>1$ low-temperature case, the total number of particles and energy can be obtained as
\begin{equation}
    \label{1}
    N=\frac{V}{3\pi^2}(8\pi^2m)^{3/2}\mu^{3/2}\left[1-\frac{3}{8}I^{q,p}\left(\frac{T}{\mu}\right)^2\right]
\end{equation}
and
\begin{equation}
    \label{2}
    U=\frac{V}{5\pi^2}(8\pi^2m)^{3/2}\mu^{5/2}\left[1-\frac{15}{8}I^{q,p}\left(\frac{T}{\mu}\right)^2\right]
\end{equation}
where the $q,p$-deformed integral $I^{q,p}$ is given by
\begin{equation}
    \label{3}
    I^{q,p}=\int^{\infty}_{0}{(\epsilon-\mu)^2\frac{\partial \eta(q,p)}{\partial \epsilon}d\epsilon}
\end{equation}
Moreover, the pressure of the system is obtained from the equation of state (EOS) $U=(3/2)PV$ for deformed fermions, such as
\begin{equation}
    \label{5}
    P=\frac{2}{3}\varepsilon=\frac{2}{15\pi^2}(8\pi^2m)^{3/2}\mu^{5/2}\left[1-\frac{15}{8}I^{q,p}\left(\frac{T}{\mu}\right)^2\right]
\end{equation}
where $\varepsilon=U/V$ is the energy density. Moreover, we know from the thermodynamical relation that $P=-\Omega$ where $\Omega$ is grand potential whose minimum value corresponds to the equilibrium state. In order to investigate the core of Q-Stars in terms of $q,p$-fermions we need to find the zero temperature thermodynamical functions of the system. In the $T=0$ limit, Eq.(\ref{1}) turns out to be
\begin{equation}
    \label{6}
    n_T(0)=\frac{(8\pi^2m)^{3/2}}{3\pi^2}\mu_0^{3/2}
\end{equation}
where $n_T=N/V$ is the total number density. Also, one can find the chemical potential by solving Eq.(\ref{1}) by the series expansion of the square bracket term and the use of Eq.(\ref{6}) as
\begin{equation}
    \label{7}
    \mu=\mu_0\left[1+\frac{1}{4}I^{q,p}\left(\frac{T}{\mu_0}\right)^2\right]
\end{equation}
One can obtain the energy density and pressure in terms of zero temperature chemical potential as
\begin{equation}
    \label{8}
    \varepsilon=\frac{3}{5}n_T(0)\mu_0\left[1-\frac{5}{4}I^{q,p}\left(\frac{T}{\mu_0}\right)^2\right]
\end{equation}
and
\begin{equation}
    \label{9}
    P=\frac{2}{5}n_T(0)\mu_0\left[1-\frac{5}{4}I^{q,p}\left(\frac{T}{\mu_0}\right)^2\right] \\.
\end{equation}

\begin{figure}
\includegraphics[width=0.48\textwidth]{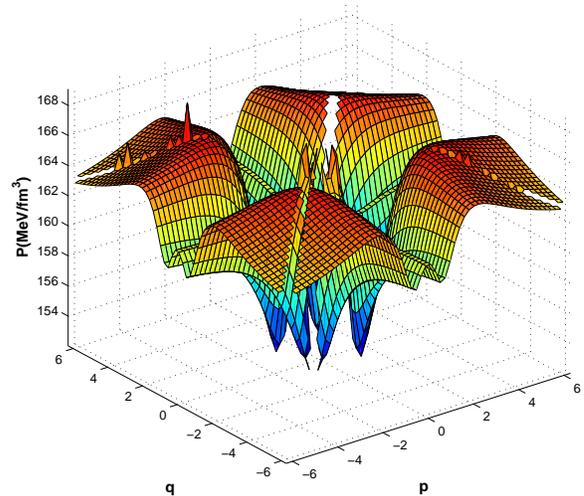}
\caption{\label{fig1}Behavior of pressure with $q,p$ deformation parameters at temperature $T=40\ \mathrm{MeV}$}
\end{figure}

\begin{figure}
\includegraphics[width=0.48\textwidth]{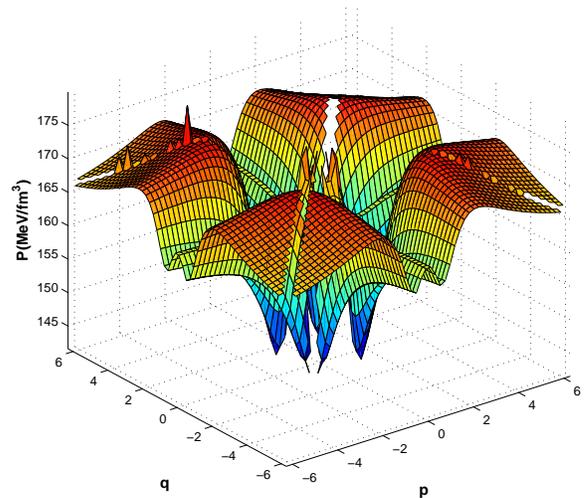}
\caption{\label{fig2}Behavior of pressure with $q,p$ deformation parameters at temperature $T=60\ \mathrm{MeV}$}
\end{figure}

In the core of compact stars, such as neutron stars, the density is generally taken to be at least three times larger than the nuclear density $n_0=0.122 \ \mathrm{fm}^{-3}$ \cite{44}. Here, we investigate the behavior of the Q-Stars for densities up to 10 times the nuclear density, and also for a particular value of 6 nuclear density. Moreover, in compact star studies, such as quark stars, the mass of the constituents is taken to be around $2-3 \ \mathrm{MeV}$ \cite{44,45}. Because the Q-Stars are more compact objects than the quark stars, we consider the mass of constituent deformed fermions as $7 \ \mathrm{MeV}$. For the total number density, Eq.(\ref{6}) becomes at $T=0$
\begin{eqnarray}
    \label{10}
    n_T(0)&=&\frac{(8\pi^{2}7)^{\frac{3}{2}}
\mu_0^{3/2}}{3\pi^{2}}1.3015\times10^{-7}\ \mathrm{fm}^{-3}  \nonumber\\
&=&5.7118\times10^{-5}\mu_0^{3/2}\ \mathrm{fm}^{-3}=0.122n \ \mathrm{fm}^{-3}
\end{eqnarray}
where the unit conversion $\mathrm{GeV}^3=1.3015 \times 10^{2}\ \mathrm{fm}^{-3}$ is used since natural units are used $h=k_B=1$. Moreover, the unit of chemical potential is taken in $\mathrm{MeV}$. The pressure can be obtained at $T=0$ from Eq.(\ref{9})
\begin{equation}
    \label{11}
    P(0)= 2.2847\times10^{-5}\mu_0^{5/2}\ \mathrm{MeV/fm}^3 \\.
\end{equation}
The value of $\mu_0$ in terms of the number factor $n$ of nuclear density can be obtained from Eq.(\ref{10}) as
\begin{equation}
    \label{12}
    \mu_0=165.86\ n^{2/3} \\,
\end{equation}
then the pressure becomes
\begin{equation}
    \label{13}
    P(0)=8.0944\ n^{5/3}\ \mathrm{MeV/fm}^3 \\.
\end{equation}
Then, we obtain the pressure (\ref{9}) in terms of the number $n$ factor of nuclear density at finite temperatures by using Eq.(\ref{12})
\begin{equation}
    \label{14}
    P=8.0944n^{5/3}\left[1-4.5439\times10^{-5}I^{q,p}n^{-4/3}T^2\right]
\end{equation}
For a particular $n=6$ nuclear density of the Q-Star, the behavior of the pressure with $q,p$ deformation parameters is illustrated in Figs.~\ref{fig1} and \ref{fig2} for temperatures $T=40\ \mathrm{MeV}$ and $T=60\ \mathrm{MeV}$, respectively. The maximum pressure which is equivalently the minimum grand potential value as the equilibrium state is obtained at $q=0.5325$ and $p=0.3195$ for deformation parameters. For $T=40\ \mathrm{MeV}$, $P=165.23\ \mathrm{MeV/fm}^3$, and for $T=60\ \mathrm{MeV}$,$P=171.78\ \mathrm{MeV/fm}^3$ are obtained.


\begin{figure}
\includegraphics[width=0.48\textwidth]{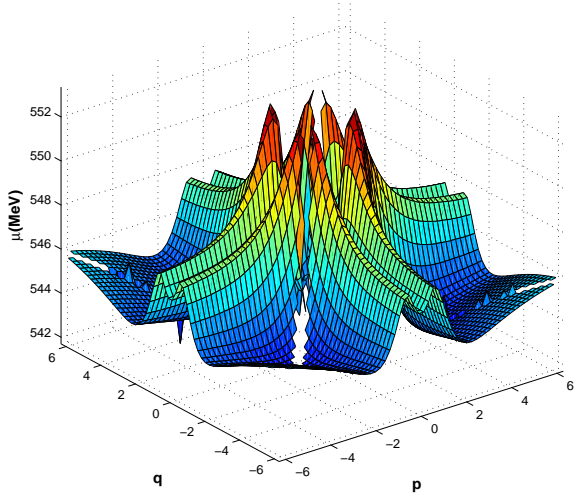}
\caption{\label{fig3}Behavior of chemical potential with $q,p$ deformation parameters at temperature $T=40\ \mathrm{MeV}$}
\end{figure}

\begin{figure}
\includegraphics[width=0.48\textwidth]{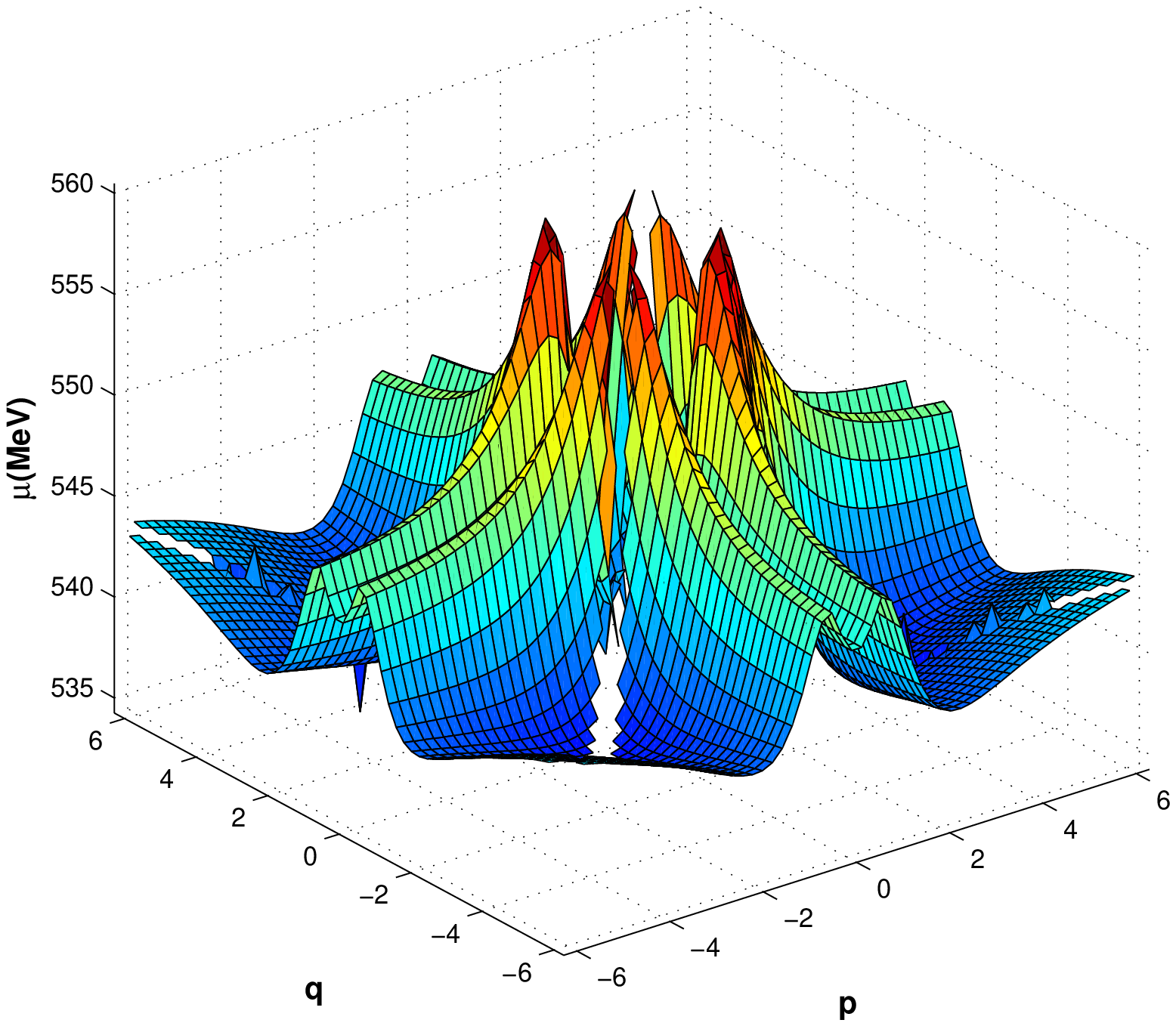}
\caption{\label{fig4}Behavior of chemical potential with $q,p$ deformation parameters at temperature $T=60\ \mathrm{MeV}$}
\end{figure}

We can also find the chemical potential (\ref{7}) at finite temperatures by using Eq.(\ref{12})
\begin{equation}
    \label{15}
    \mu=165.86\ n^{2/3}\left[1+9.0878\times10^{-6}I^{q,p}n^{-4/3}T^2\right] \\.
\end{equation}
The behavior of the chemical potential is represented in Figs.~\ref{fig3} and \ref{fig4} for temperatures $T=40\ \mathrm{MeV}$ and $T=60\ \mathrm{MeV}$, respectively. $q=0.5325$ and $p=0.3195$ values of deformation parameters giving the maximum pressure yields the chemical potential, at $T=40\ \mathrm{MeV}$, $\mu=544.09\ \mathrm{MeV}$, and at $T=60\ \mathrm{MeV}$, $\mu=539.63\ \mathrm{MeV}$ are obtained.


\begin{figure}
\includegraphics[width=0.48\textwidth]{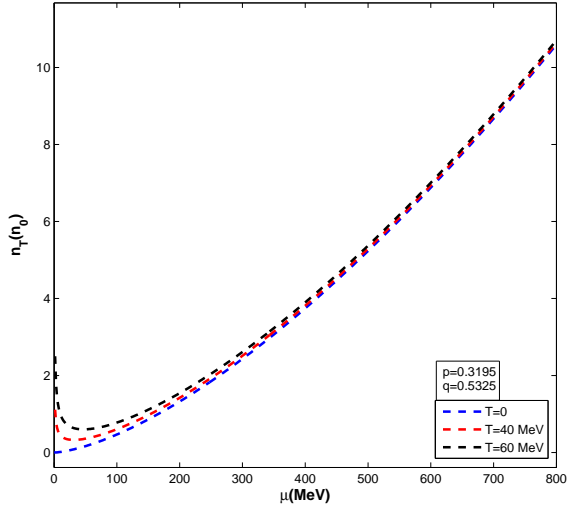}
\caption{\label{fig5}Dependence of total number density on chemical potential for temperatures $T=0$, $T=40\ \mathrm{MeV}$, $T=60\ \mathrm{MeV}$}
\end{figure}

Moreover, we can find the total number density in finite temperatures by using unit conversions in Eq.(\ref{1}) as follows
\begin{eqnarray}
    \label{16}
    n_T&=&\frac{(8\pi^{2}7)^{\frac{3}{2}}
\mu^{3/2}}{3\pi^{2}}1.3015\times10^{-7}\mathrm{fm}^{-3} \times \nonumber \\ 
&&\qquad \qquad \qquad \qquad \qquad \left[ 1-\frac{3}{8}I^{q,p}\left( \frac{T}{\mu}\right)^{2}\right] \nonumber \\
&=&5.7118\times10^{-5}\mu^{3/2}\left[ 1-\frac{3}{8}I^{q,p}\left( \frac{T}{\mu}\right)^{2}\right]\ \mathrm{fm}^{-3} \qquad 
\end{eqnarray}
The dependence of total number density on chemical potential is illustrated in Fig.\ref{fig5} for temperatures $T=0$, $T=40\ \mathrm{MeV}$ and $T=60\ \mathrm{MeV}$. After a certain chemical potential value about $\mu=400\ \mathrm{MeV}$, the temperature has no effect on the dependency between number density and chemical potential. In addition, the higher values of chemical potential yield increasing number densities independent from the temperature. 

\begin{figure}
\includegraphics[width=0.48\textwidth]{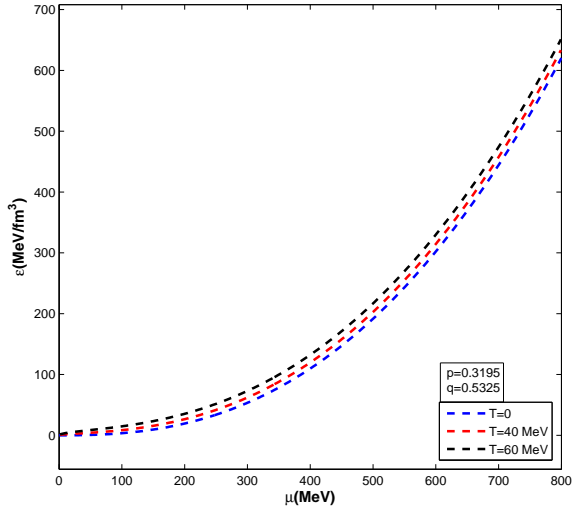}
\caption{\label{fig6}Dependence of energy density on chemical potential for temperatures $T=0$, $T=40\ \mathrm{MeV}$, $T=60\ \mathrm{MeV}$}
\end{figure}

\begin{figure}
\includegraphics[width=0.48\textwidth]{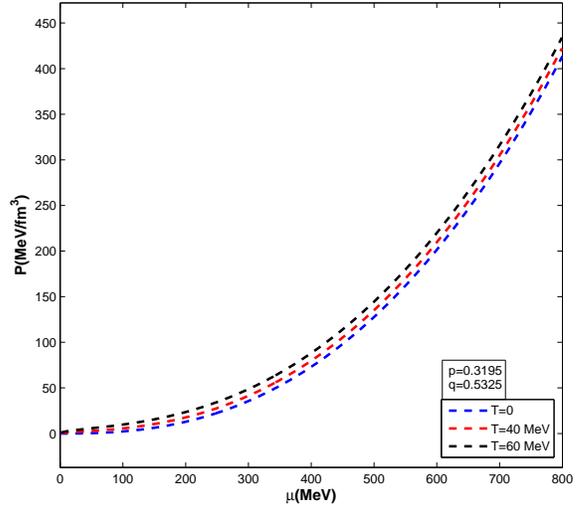}
\caption{\label{fig7}Dependence of pressure on chemical potential for temperatures $T=0$, $T=40\ \mathrm{MeV}$, $T=60\ \mathrm{MeV}$}
\end{figure}

Similarly, we can obtain the energy density and the pressure in finite temperatures in Eq.(\ref{2}) and Eq.(\ref{5}) as
\begin{eqnarray}
    \label{17}
    \varepsilon&=&\frac{(8\pi^{2}7)^{\frac{3}{2}}
\mu^{5/2}}{5\pi^{2}}1.3015\times10^{-7}\mathrm{MeV.fm}^{-3} \times \nonumber \\ 
&&\qquad \qquad \qquad \qquad \qquad \left[ 1-\frac{15}{8}I^{q,p}\left( \frac{T}{\mu}\right)^{2}\right] \nonumber \\
&=&3.4269\times10^{-5}\mu^{5/2}\left[ 1-\frac{15}{8}I^{q,p}\left( \frac{T}{\mu}\right)^{2}\right]\ \mathrm{MeV.fm}^{-3} \nonumber\\
 &&\quad
\end{eqnarray}
and
\begin{equation}
    \label{18}
    P=2.2846\times10^{-5}\mu^{5/2}\left[ 1-\frac{15}{8}I^{q,p}\left( \frac{T}{\mu}\right)^{2}\right]\ \mathrm{MeV.fm}^{-3}
\end{equation}
The behavior of energy density and the pressure with respect to the chemical potential for temperatures $T=0$, $T=40\ \mathrm{MeV}$, $T=60\ \mathrm{MeV}$ are given in Figs.~\ref{fig6} and \ref{fig7}, respectively. In these figures, higher temperatures gives higher pressure and energy density values. Moreover, the energy density and pressure values show a dramatic increase after a chemical potential value around $\mu=300\ \mathrm{MeV}$. In Figs.~\ref{fig5}-\ref{fig7}, we infer that the increasing values of chemical potential $\mu$ gives increasing values of number density $n_T$, energy density $\varepsilon$ and pressure $P$ values of the Q-Star system, respectively.


\begin{figure}
\includegraphics[width=0.48\textwidth]{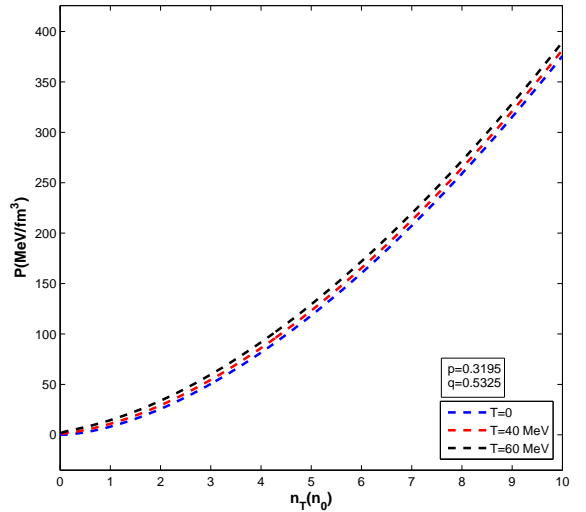}
\caption{\label{fig8}Behavior of pressure with number density for different temperatures $T=0$, $T=40\ \mathrm{MeV}$, $T=60\ \mathrm{MeV}$}
\end{figure}

\begin{figure}
\includegraphics[width=0.48\textwidth]{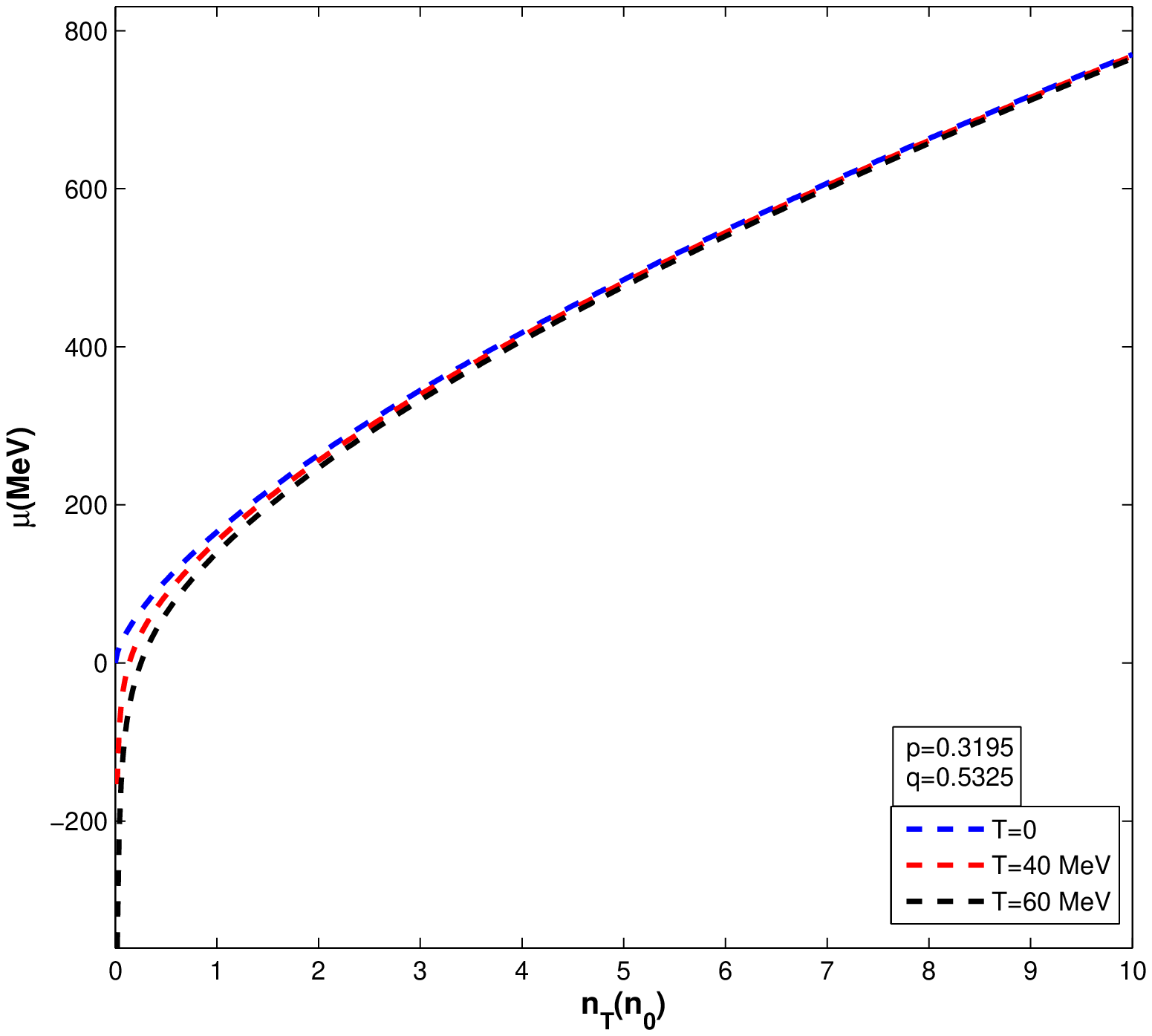}
\caption{\label{fig9}Behavior of chemical potential with number density for different  temperatures $T=0$, $T=40\ \mathrm{MeV}$, $T=60\ \mathrm{MeV}$}
\end{figure}

Moreover, the behavior of pressure in Eq.(\ref{14}) and chemical potential in Eq.(\ref{15}) with respect to the number density for temperatures $T=0$, $T=40\ \mathrm{MeV}$, $T=60\ \mathrm{MeV}$ are represented in Figs.~\ref{fig8} and \ref{fig9}. According to these figures, when the number density $n_T$ increases, pressure and chemical potential increases for the system. On the other hand, we see that there occurs a high particle production in high temperature regimes, although the chemical potential is zero, or even less than zero due to the temperature, in Figs.~\ref{fig5} or \ref{fig9}.

\section{\label{sec3}Configurations of Hot Q-Stars}

We investigate the possible configurations of formation and evolution of cold and hot Q-Stars in temperatures of $T=0$, $T=40\ \mathrm{MeV}$ and $T=60\ \mathrm{MeV}$. In the literature, Q-Stars are expected to have a maximum radius about two times greater than the Schwarzschild radius meaning it is very close to the mass-radius relations of the black holes. However, they are not yet dense as much as to form a black hole, therefore the gray hole term is used to express Q-Stars which enables light to escape from its gravitational field. The constituent of Q-Stars are expected to be SUSY particles, or Q-Balls with a large particle density. Here we investigate the  Q-Stars with $q,p$-fermions constituent.

\begin{figure}
\includegraphics[width=0.48\textwidth]{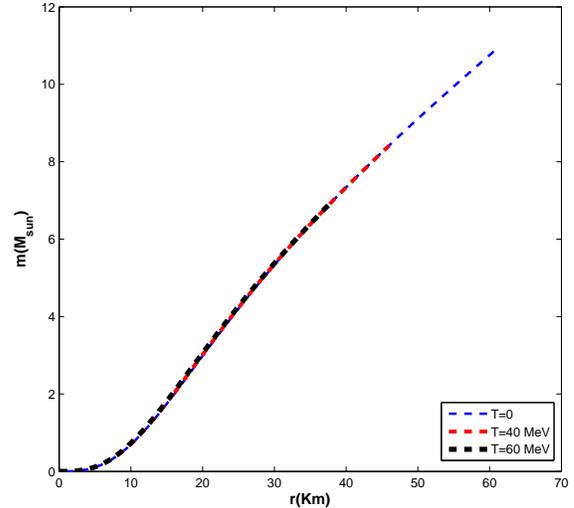}
\caption{\label{fig10}Solution of TOV equation for accumulated mass inside the Q-Star with central density $n_T(r=0)=6n_0$ for temperatures $T=0$, $T=40\ \mathrm{MeV}$, $T=60\ \mathrm{MeV}$}
\end{figure}

\begin{figure}
\includegraphics[width=0.48\textwidth]{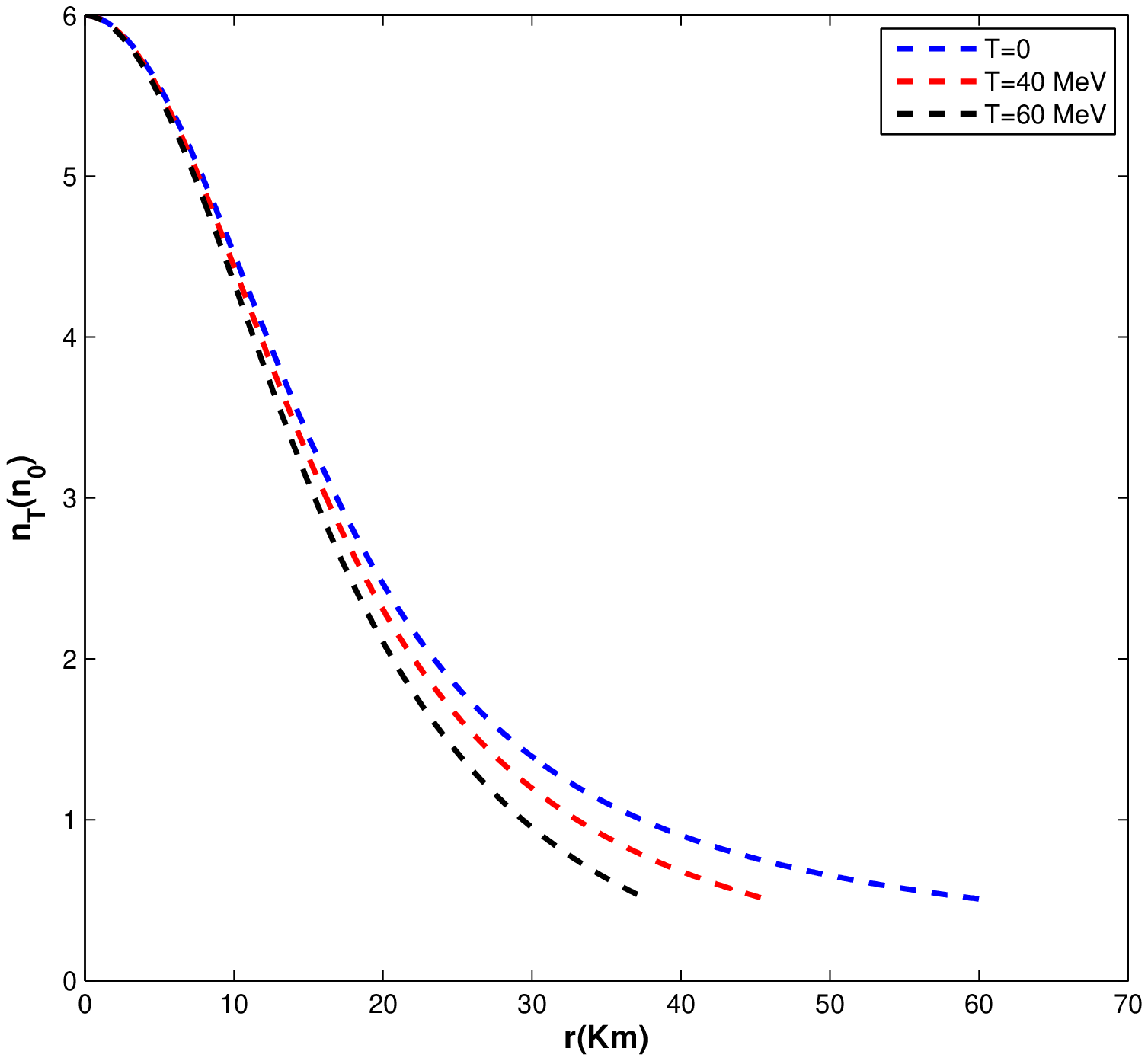}
\caption{\label{fig11}Solution of TOV equation for matter distribution inside the Q-Star with central density $n_T(r=0)=6n_0$ for temperatures $T=0$, $T=40\ \mathrm{MeV}$, $T=60\ \mathrm{MeV}$}
\end{figure}

\begin{figure}
\includegraphics[width=0.48\textwidth]{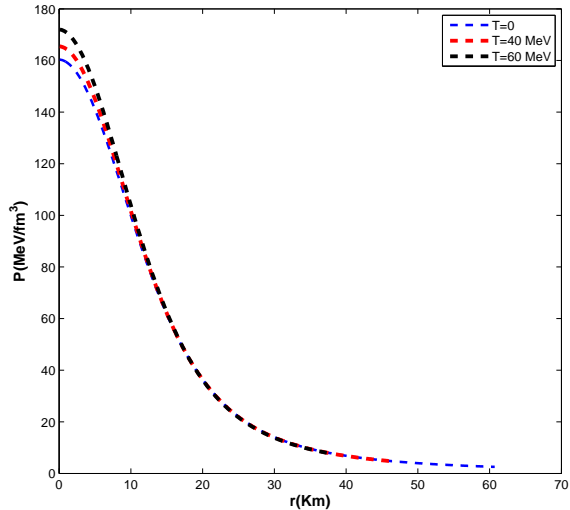}
\caption{\label{fig12}Solution of TOV equation for pressure inside the Q-Star with central density $n_T(r=0)=6n_0$ for temperatures $T=0$, $T=40\ \mathrm{MeV}$, $T=60\ \mathrm{MeV}$}
\end{figure}

The Tolman–Oppenheimer–Volkoff (TOV) equations for spherically symmetric configurations stars are applied to Q-Stars with $q,p$-deformed fermions \cite{46},
\begin{equation}
    \label{19}
    \frac{dP(r)}{dr}=\frac{\left[\varepsilon(r)+P(r)\right]\left[m(r)+4\pi r^3P(r)\right]}{r\left[r-2m(r)\right]}
\end{equation}
where $\varepsilon(r)$ and $P(r)$ are the energy density and pressure values at a distance $r$ from the center of the star. The mass $m(r)$ in the spherical radius of $r$ is given by
\begin{equation}
    \label{20}
    m(r)=\int_0^r{\varepsilon(r')4\pi r'^2dr'}
\end{equation}
In order to solve these equation, it is necessary to define the boundary conditions at the center of the star. The mass is considered as zero, the pressure and number density $n_T$ are maximum at the center. When the star reaches the maximum possible radius its mass becomes maximum, however the pressure and number density reaches almost a zero value. According to the thermodynamical relations in previous section, we first choose an $n=6$ nuclear density initial value to investigate the dependence of the mass and the central number density as well as the pressure on the radius for different temperatures $T = 0$, $T=40\ \mathrm{MeV}$, $T=60 \ \mathrm{MeV}$ defining a configuration. We used the cut off value for the number density $n_T=0.5n_0$ because the model below this cut off density gives unphysical low-density quasi-free deformed fermions. We illustrate the results in Figs.~\ref{fig10}-\ref{fig12}. 

\begin{figure}
\includegraphics[width=0.48\textwidth]{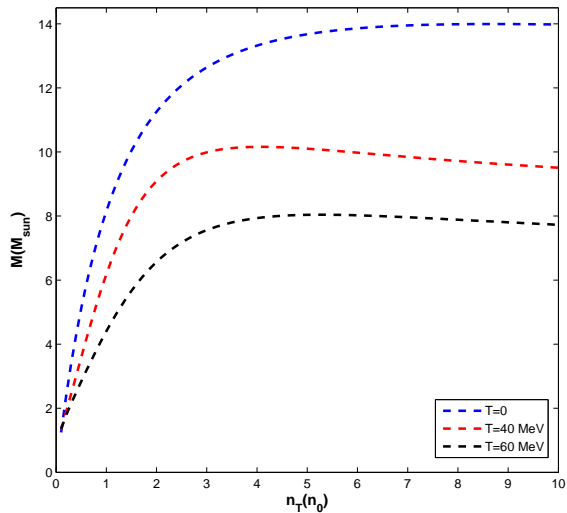}
\caption{\label{fig13}Stable configurations of Q-Stars for different temperatures $T=0$, $T=40\ \mathrm{MeV}$, $T=60\ \mathrm{MeV}$ with total mass as a function of
central number density}
\end{figure}

In Fig.~\ref{fig10}, it is clear that higher temperatures lead to the less massive and smaller size Q-Stars. We can easily see from Fig.~\ref{fig11} that temperature has no role on the mechanical equilibrium of Q-Star core around the center of the configuration because of its high density. The effect of temperature is significant on the outer peripheral regions and changes the total mass of the star for a certain central number density, as in Figs.~\ref{fig10} and \ref{fig13}.

In addition, the temperature leads to the increase of pressure but this gives a quick fall off the pressure of Q-Star (see Fig.~\ref{fig12}) and yields a smaller $M-R$ relation. When the temperature increases in Fig.~\ref{fig11}, the number density $n_T$ decreases, and then a less denser and smaller size Q-Star exists.


\begin{figure}
\includegraphics[width=0.48\textwidth]{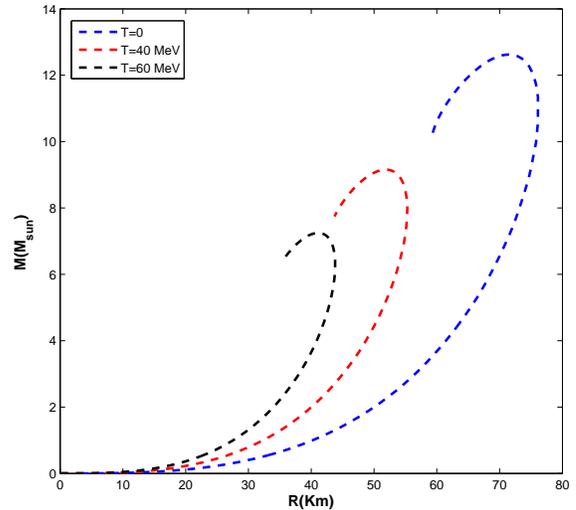}
\caption{\label{fig14}Stable configurations of Q-Stars for different temperatures $T=0$, $T=40\ \mathrm{MeV}$, $T=60\ \mathrm{MeV}$ with total mass as a function of
maximum radius}
\end{figure}

By using the findings of above configurations, we can obtain the total radius of the Q-Star core $R$ for different number density values ranging to 10 central nuclear density. Moreover, the total mass $M(R)$ can also be obtained for these varying number density values up to $10n_0$. The stable configurations of Q-Stars with maximum mass at the maximum radius for different temperatures $T=0$, $T=40\ \mathrm{MeV}$, $T=60\ \mathrm{MeV}$ are given in Figs.~\ref{fig13} and \ref{fig14} such that the mass as a function of central number density and of the rotal radius, respectively.

Fig.~\ref{fig13} shows increasing temperatures gives Q-Star configurations with smaller masses. When the star gets cooler we obtain more massive Q-Stars for a same nuclear density value. This fact is also clear from the Fig.~\ref{fig14} that the higher temperatures gives smaller Q-Star configurations in total mass and total radius. This implies that the temperature rises the pressure and lowers the density of the star, which yields a smaller mass and a smaller radius Q-Star configuration. Moreover, there exist some Q-Star configurations having two different radii values for a single mass value between $R=[60-75] \ \mathrm{Km}$ values for $T=0$ case, between $R=[40-50] \ \mathrm{Km}$ values for $T=40\ \mathrm{MeV}$ case and $R=[30-40] \ \mathrm{Km}$ values for $T=60 \ \mathrm{MeV}$ case. Similar case is also valid for the Q-Stars with one same radius and having different total masses at this radius for the same regions of three temperature values.


\section{Conclusions}

In this study, we have investigated the effect of interpolating property of deformation parameters on the thermodynamics
of $q,p$-deformed fermions under the conditions of maximum pressure and equilibrium suitable for
the discussion of Q-Stars. The EOS and related thermodynamical quantities have been derived for the $q,p$-deformed fermion system as the constituents of the Q-Star structure. The deformed model
parameters are chosen such that the pressure becomes maximum, or equivalently the grand potential becomes minimum implying the equilibrium state. It is obtained that the increase of temperature leads to a higher average pressure and chemical potential values in Figs.~\ref{fig1}-\ref{fig4}.

The dependence of thermodynamical functions such as total number density, energy density and the pressure on chemical potential are illustrated in Figs.\ref{fig5}-\ref{fig7} for temperatures $T=0$, $T=40\ \mathrm{MeV}$ and $T=60\ \mathrm{MeV}$. We infer from Fig.~\ref{fig5} that the particles can still be produced for zero, or even less than zero chemical potential values for hot Q-Star profiles. However, after a certain chemical potential value around $\mu=400\ \mathrm{MeV}$, the temperature has no effect on the dependency of particle number density. The behavior of energy density and the pressure is such that the higher temperatures gives higher pressure and energy density values with a dramatic increase after a chemical potential value of $\mu=300\ \mathrm{MeV}$.

After analyzing the thermodynamical relations and EOS of the Q-Star profiles with $q,p$-fermion constituents, we investigate the evolution of star interior with TOV equations. As a first sample profile, we consider a $6n_0$ nuclear density central number density for cold and hot star configurations at $T = 0$ and $T=40\ \mathrm{MeV}$, $T=60 \ \mathrm{MeV}$, respectively. We show the results in Figs.~\ref{fig10}-\ref{fig12} according to which the higher temperatures lead to the less massive and smaller size Q-Stars. We also infer that the temperature has no role on the mechanical equilibrium of Q-Star core around the center of the configuration because of its high density. On the other hand, the temperature has an effect on the outer regions and changes the total mass of the star. Moreover, when the temperature  the increases pressure of the star increases but this is led by a quick fall off the pressure of Q-Star as in Fig.~\ref{fig12}, and also by a smaller $M-R$ relation. In addition, the increase of temperature leads to the decrease in number density and a less denser and smaller size Q-Star, as it is clear from the Figs.~\ref{fig11} and \ref{fig12} for the pressure and number density versus radius, respectively. The pressure remains constant for radius values greater than $10\ \mathrm{Km}$ and the number density of the Q-Star decreases for higher temperatures, which leads to less denser star profiles for hotter star configurations.

In order to investigate the stable configurations, we also obtain the total mass and radius relations of the Q-Star by changing the number density values up to 10 central nuclear density in the TOV equations. Figs.~\ref{fig13} and \ref{fig14} illustrate the dependence of total mass and the total radius on the central number density, respectively. In Fig.~\ref{fig13}, we see the effect of higher temperatures as giving smaller mass values for Q-Star configuration. When the star gets cooler we obtain more massive Q-Stars for a same nuclear density value. This fact is also obvious from the Fig.~\ref{fig14} that the higher temperatures gives smaller Q-Star configurations in total mass and total radius. Moreover, there exist some Q-Star configurations having two different radii values for a single mass value between $R=[60-75] \ \mathrm{Km}$ values for $T=0$ case, between $R=[40-50] \ \mathrm{Km}$ values for $T=40\ \mathrm{MeV}$ case and $R=[30-40] \ \mathrm{Km}$ values for $T=60 \ \mathrm{MeV}$ case. Similar case is also valid for the Q-Stars with same radius values but having different total masses at this radius for three different temperature cases.

As a concluding remark, we could obtain the Q-Star structures, also called as gray holes having a radius of two times greater than the corresponding Schwarzschild radius by considering the constituents as a generalizations of $q$-deformed fermions by two deformation parameters $q$ and $p$. Instead of defining the interactions between the fermions by complicated Lagrangian function, we choose to consider $q,p$-deformed fermions in order to control the interactions by deformation parameters such that the maximum Q-Star pressure, minimum grand potential are obtained for the equilibrium configuration of the star. The most important result of the study is the existence of Q-Star configuration that can reach up to $13$ solar mass and $75$ Km radius for these $q,p$-fermion constituents, which is consistent with the general Q-Star structure in the literature. Moreover, it is found that when the star gets cool down from high temperatures the total mass and radius of the star increases, and a more denser state is reached. 

\section*{Acknowledgements}
The authors are grateful to Elif Dil, PhD at Hacettepe University, Statistics Department, for writing the matlab codes for numerical calculations.

\bibliography{apssamp}

\end{document}